\documentstyle[preprint,aps]{revtex} 
\begin{document}
\draft

\title{Density-Functional Theory of Polar Insulators}
\author{X. Gonze$^{\star}$, Ph. Ghosez$^{\star}$, and R.W.
Godby$^{\dagger}$}
\address{$^{\star}$ Unit\'e PCPM,
Universit\'e Catholique de Louvain,
B-1348 Louvain-la-Neuve, Belgium}
\address{$^{\dagger}$ Department of Physics, University of York,
Heslington, York YO1 5DD, U.K.}
\date{\today}
\maketitle
\begin{abstract}
We examine the density-functional theory of macroscopic insulators,
obtained in the large-cluster limit or under periodic boundary conditions.
For {\it polar} crystals, we find that the two procedures are not equivalent.
In a large-cluster case, the exact exchange-correlation potential acquires
a homogeneous ``electric field'' which is absent from the usual local
approximations, and the Kohn-Sham electronic system becomes metallic.
With periodic boundary conditions, such a field is forbidden,
and the polarization deduced from Kohn-Sham wavefunctions is incorrect
even if the exact functional is used.
\end{abstract}
%


\newpage

Although the density-functional theory (DFT) introduced by Hohenberg, Kohn
and Sham \cite{Hohenberg64,Kohn65}
has become the standard method for first-principles calculations of
the ground-state properties of solids, to our
knowledge~\cite{DFTReviews}, the implications of applying DFT to
infinite, insulating, crystals have not been fully
appreciated. In part, this reflects the fact that the key theorems
of DFT \cite{Hohenberg64,Kohn65} were proved for arbitrarily large,
but not infinite, systems. In the present paper, we show that
the exact DFT treatment of {\it polar} crystals (a) with
the usual Born-von Karman (BvK) boundary conditions,
or (b) from the macroscopic limit of large
clusters, will generally give different macroscopic polarizations.
Only (b) is correct.

Investigating the {\it response} of periodic insulators to a homogeneous
electric field, we recently revealed~\cite{Gonze95}
the polarization-dependence of the exchange-correlation energy, and
its consequences on the dielectric response.
Aulbur, J\"onsson and Wilkins~\cite{Aulbur96} quantified this effect
for real materials, while Resta~\cite{Resta96} discussed the origin of
such a behaviour in connection with long-range correlation effects.
The present study emphasizes a more basic role of the polarization
in DFT: careful handling of the polarization is mandatory for
polar solids, even under zero electric field.
In polar materials, the spontaneous polarization computed from the
Kohn-Sham (KS) wavefunctions will be correct only if
an exchange-correlation homogeneous electric field is allowed
throughout the material, in which case the KS electronic
system becomes {\it metallic}. This field will appear in the exact DFT
treatment of a finite cluster but is forbidden when using BvK conditions.
Approximate density-functionals such as the Local Density Approximation (LDA)
and the Generalized Gradient Approximation (GGA) always fail to
yield an exchange-correlation electric field: within these approximations,
using BvK boundary conditions or finite clusters incorrectly provide
the {\it same} value of the
polarization~\cite{LDAGGA}. Any improvement to these functionals which
retains a dependence only on the periodic density will be similarly
flawed. We will exhibit our results for a one-dimensional
model semiconductor.

The correct definition of a macroscopic crystal
is clearly as the limit
of a finite crystal of increasing size.
Fig. 1(a) shows schematically the total electrostatic potential
$V_{\rm elec} = V_{\rm e} + V_{\rm N} +
 V_{\rm appl}$ in such a finite crystal,
where $V_{\rm e}$
is the electrostatic potential due to the
ground-state electron density, $V_{\rm N}$ is the potential
due to the nuclei,  and $V_{\rm appl}$ is an applied potential,
created by an external short-circuited capacitor,
that maintains equality of the electrostatic potential
on the two sides~\cite{capacitor}.
The sum $V_{\rm ext} = V_{\rm N} + V_{\rm appl}$
is referred to as the external
potential. The total electrostatic
potential in the bulk region is periodic and, crucially for a
non-zero polarization, non-centrosymmetric. The potential
just outside the surface is fixed by the electrostatic potential of
the capacitor plates. The corresponding
ground-state electron density is also shown.
In the bulk region, it is periodic, with the
same periodicity as the local potential~\cite{CDW}. Close
to the surface, the density deviates from
perfect periodicity, although this effect decreases exponentially
with the distance from the surface~\cite{Kohn64}.

The macroscopic polarization of such a finite solid is directly linked to
the total surface charge~\cite{Landau60,Vanderbilt93}.
Its value is equal to zero (modulo a half-quantum) if the crystal is
centrosymmetric
\cite{Vanderbilt93,Resta94}, but otherwise can have any value and must
be calculated.
For a long time the macroscopic polarization was only accessible from
the surface charge and was a well-defined concept only for finite solids.
Recent theoretical advances have shown that it can also
conveniently be determined, up to a quantum, from a Berry phase of the
correlated many-body wavefunction of the {\it bulk}
~\cite{Vanderbilt93,Resta94,King-Smith93,Ortiz94,Resta95}.
Within this approach the macroscopic polarization appears as a bulk property
and is unambiguously defined even for the infinite periodic solid,
which is of practical interest in solid state {\it ab initio} calculations.

The breakthrough of King-Smith and Vanderbilt~\cite{King-Smith93},
leading to the modern theory of the polarization, was actually
carried out in the context of DFT. Later,
they argued~\cite{Vanderbilt93} that
the Berry phase of the occupied KS wavefunctions possesses
an exact physical meaning
since the surface charge must be exactly reproduced within DFT
\cite{Vanderbilt93}. We now show that the justification of Vanderbilt
and King-Smith apply to exact DFT {\it only} when considering
{\it finite} solids, and not when applying BvK periodic boundary conditions.

In the context of DFT, it is shown that the density $n({\bf r})$ of the
ground state of a system uniquely determines the external potential
up to a constant. Following Kohn and Sham~\cite{Kohn65}, one can
introduce a fictitious system of non-interacting electrons in
an effective potential $V_{\rm eff}=V_{\rm ext}+V_{\rm H}+V_{\rm xc}$
(where $V_{\rm xc}$ is the exchange correlation potential)
that reproduces the ground-state electron density of the real system.
In the particular case where {\it periodic boundary conditions} are imposed,
although the KS effective potential $V_{\rm eff}$ is constrained to reproduce
the correct {\it periodic} bulk density of the polar solid, there is no
guarantee
that it will reproduce the correct {\it polarization}, since this information
is not contained in the criterion for the effective potential to be
correct~\cite{notemetal}. Such a periodic DFT is based only on the periodic
part of the density, while the polarization is a completely independent
quantity \cite{Resta94,Gonze95,Resta96} that depends
on the phase of the correlated
wavefunctions. The polarization will be correct only in those solids
where a fundamental symmetry (such as centrosymmetry)
constrains the polarization, or where external parameters,
such as the pressure, are fortuitously chosen.

We illustrate this for a one-dimensional model semiconductor
~\cite{footnote1}.
In this model the electrostatic potential
is periodic and asymmetric:
$  V_{\rm elec}(x)=
            V_{c} \,  \cos  \frac{2 \pi x}{a}
          + V_{s} \, \sin  \frac{4 \pi x}{a}
$.
A non-local self-energy operator, intended to mimic the relevant
many-body effects, has the same non-local form as in
Ref.~\cite{Gonze95}:
$ 
 \Sigma (x,x',\omega)=
  \frac {f(x)+f(x')}{2} g(|x-x'|)
$ 
where $f(x)=F_{o} [1-\cos\frac{2 \pi x}{a}]$
is a negative function
with the periodicity of one unit cell and $g(y)$ is a normalized
gaussian of width $w$.

First, the Schr\"odinger equation containing the self-energy
operator is solved by direct diagonalization using a plane-wave basis set.
The density is deduced
from the sum of the squares of the
eigenfunctions. From this result, using standard iterative
optimization techniques, we construct an {\em exact} density-functional
theory by determining the local potential $V_{\rm eff}(x)$
which, when filled with {\em non-interacting} electrons
(no self-energy operator), reproduces the same electron density
as in the self-energy calculation.
Fig. 2 presents the function $V_{\rm elec}(x)$, as well as the
density $n(x)$,
and the effective potential $V_{\rm eff}(x)$,
for the following set of parameters :
$a_0= 4$ a.u., $V_c=V_s=2.72$ eV, $F_o=-4.08$ eV, $w=2$ a.u.

Using the Berry-phase approach~\cite{Resta94,King-Smith93},
we then compute the polarization~\cite{footnote2}.
In the self-energy calculation, the polarization is
$22.68 \,\, 10^{-3}$ electrons with respect
to the centrosymmetric system with $V_s=0$, while that calculated
from the Berry phase of the Kohn-Sham wave-functions
is $21.99 \,\, 10^{-3}$ electrons. The two polarizations
differ by $3$\%, well outside the calculational error bar.
This value may be taken as an order of magnitude
estimation of the effect in real materials,
and is compatible with the observed (and often satisfactory) accuracy
of LDA polarization calculations for real ferroelectric materials
\cite{Resta93}.

The deficiency in the periodic-boundary approach,
reflected in the Berry phase of the KS wavefunctions
and hence in the polarization, is that the
exchange-correlation potential is prevented from having a component
which is linear in space \cite{periodic}. The KS theorem demonstrates
that there is only one periodic effective potential $V_{\rm eff}$
that reproduces a particular periodic density. However, once an additional
linear component is allowed, there exists an infinite family
of KS potentials that gives the same periodic density but different
polarizations \cite{Gonze95}. Imposing BvK conditions on the potential
thus arbitrarily constrains the polarization to a specific,
usually incorrect, value.
This restriction does not apply for the finite cluster, where
application of the KS theorem shows that there exists a unique
effective potential that, when used in the effective Hamiltonian,
will generate the exact ground-state density {\it everywhere}:
not only in the bulk
region (as in the BvK case), but {\it also} in the surface region,
resulting in the correct polarization. Fig. 1(c) sketches the behaviour
of such an effective potential. The linear part
is necessary to yield the correct polarization in polar crystals.
This ``exchange-correlation electric field" originates in the
ultra non-local dependence of the exchange-correlation energy in the
surface charge pointed out in Ref.~\cite{Gonze95}.

In the small cluster shown in Fig. 1(c), the magnitude of the
exchange-correlation field is approximately independent of the
cluster size, since the polarization correction relative to
periodic DFT is constant.
As the cluster is made larger, a point will be reached where the
variation in potential from one side of the cluster to the other,
due to the homogeneous exchange-correlation electric field, reaches the
DFT band gap of the material. Beyond this point, the KS electronic system is
{\it metallic} and
the band edges will ``pin" the effective potential (Fig. 3). As the cluster
is made
still larger, charge will flow freely from one face
to the other in order to maintain the
correct macroscopic polarization.
The magnitude
of the homogeneous electric field will now change with the
size of the cluster in order to maintain the potential
drop: in the
limit of large cluster size, the effective homogeneous electric field will be
infinitesimally small, although non-zero. As mentioned
in Ref.~\cite{Nunes}, an infinite
system cannot sustain a finite homogeneous electric field in its ground-state.
Here, an infinitesimal field appears naturally in the DFT treatment
of polar solids.

There is a strong similarity between this behaviour and that of a system
of two distant, different, open-shell atoms
\cite{vonBarth85}, in which the exact exchange-correlation potential
exhibits a long-range spatial variation to align the
Kohn-Sham eigenvalues. There is also a connection with
the DFT metal/insulator paradox~\cite{Godby89} in which an insulating
system may be described as metallic in DFT.

In summary, for a polar insulator, when Born-von Karman periodic
boundary conditions are used, the polarizations calculated from
the Berry phase of the Kohn-Sham wavefunctions and from the Berry
phase of the correlated wavefunction will differ, because the DFT
effective potential is prevented from acquiring a linear part.
When a large cluster is used for the DFT calculation,
a homogeneous effective exchange-correlation ``electric field''
develops in order to correctly reproduce the polarization.
The Kohn-Sham system becomes metallic.

The authors thank R. Resta, R. Martin, H. Krakauer, R.E. Cohen
and I. Mazin for interesting discussions, and
acknowledge financial support from
FNRS-Belgium (X.G.), the Concerted Actions program (X.G.),
The Royal Society and
the Engineering and Physical Sciences Research Council (R.W.G.),
the European Union (Human Capital and Mobility Program Contracts
CHRX-CT940462 and CHRX-CT930337), and the
Academic Research Collaboration Programme
between the FNRS, the CGRI, and the British Council.



\begin{center}
\section*{Figure Captions}
\end{center}

\begin{enumerate}

\item[Fig.1]
(a) The local electrostatic potential (external plus Hartree)
of an insulator, and the corresponding ground-state density. 
In the bulk region the potential is periodic.
Short-circuited capacitor plates are also present.
(b) The effective potential that, when used in Kohn-Sham
equations, is able to reproduce the periodic part of the
density shown in (a),
under Born-von Karman periodic boundary
conditions.
The macroscopic polarization is not correct.
(c) The effective potential that, when used in Kohn-Sham
equations, is able to reproduce the
density shown in (a), in all the regions of space.
The macroscopic polarization is correct (in contrast to (b)).

\item[Fig.2]
The electrostatic potential $V_{\rm elec}(x)$, the
electron density $n(x)$ and the Kohn-Sham effective
potential $V_{\rm eff}(x)$ of the model one-dimensional semiconductor
are shown when periodic boundary conditions are imposed (corresponding
to Fig. 1(b)). The Kohn-Sham electrons correctly reproduce the electron
density, but not the macroscopic polarization.

\item[Fig.3]
For sufficiently large clusters, the exchange-correlation field will
cause band overlap and hence metallization. Further increase in cluster
size leaves the band edges pinned as shown, and charge transfer occurs
between the two surfaces.

\end{enumerate}


\begin{references}
\bibitem{LDAGGA}
This can be viewed as a practical advantage of LDA or GGA:
the computation of the polarization under BvK conditions
(from a Berry phase of the KS wavefunctions) is intrinsically 
easier than from the limit of large clusters and constitutes a
consistent procedure within this approximation.
\bibitem{capacitor}
For the simplicity of the argument, we imagine that the capacitor
is a classical device, in which case the analysis of
electrostatics performed in Ref.~\cite{Resta96} applies.
\bibitem{CDW}
In this
work, we do not consider symmetry-breaking effects
like charge- or spin-density waves.
\bibitem{notemetal}
In some cases, even with BvK conditions, the Kohn-Sham system will
be a metal~\cite{Godby89}
in which case the
computation of the Berry phase is inappropriate. Henceforth
we will suppose that the KS system obtained under BvK
boundary conditions is insulating.
\bibitem{footnote2}
The standard Berry phase analysis [10-14] is applicable
to our system, showing the
exact polarization to be proportional to the Berry phase of the
eigenfunctions of the non-local Hamiltonian.
This is distinct from the Berry phase of the Kohn-Sham wavefunctions.
\bibitem{periodic}
The electrostatic potential $(V_{\rm ext}+ V_{\rm e})$
is the same periodic potential as in the many-body case,
thus constraining $V_{\rm xc}$ to be periodic also.
\end{references}
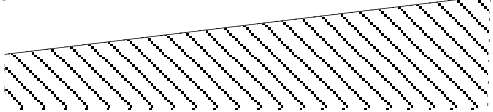
\end{document}